# SMOOTHED PARTICLE HYDRODYNAMICS METHOD FOR TWO-DIMENSIONAL STEFAN PROBLEM

Dede Tarwidi[1,2]

[1] Graduate School of Natural Science and Technology , Kanazawa University
Kakuma, Kanazawa 920-1192, Japan
[2] Faculty of Mathematics and Natural Sciences, Institut Teknologi Bandung
Jl. Ganesha 10, Bandung 40132, Indonesia
email: dede.tarwidi@yahoo.com

*Abstract*. Smoothed particle hydrodynamics (SPH) is developed for modelling of melting and solidification. Enthalpy method is used to solve heat conduction equations which involved moving interface between phases. At first, we study the melting of floating ice in the water for two-dimensional system. The ice objects are assumed as solid particles floating in fluid particles. The fluid and solid motion are governed by Navier-Stokes equation and basic rigid dynamics equation, respectively. We also propose a strategy to separate solid particles due to melting and solidification. Numerical results are obtained and plotted for several initial conditions.

Keywords: SPH, enthalpy method, Stefan problem, moving interface

## 1. INTRODUCTION

Stefan problem is related to heat transfer problem involving phase-change such as solid to liquid (melting) or liquid to solid (solidification). This issue has the attention from many researchers since the last few decades. Several methods are used to solve the Stefan problem both analytical and numerical solutions. Due to its complexity, the analytical solution is limited to one-dimensional and simple two-dimensional problem. The detail of the analytical and numerical solution of the one-dimensional Stefan problem is discussed in [1]. As to the numerical solution, in general, it can be divided into two methods, the grid/mesh-based method and the gridless/meshless-based method. Finite difference, finite element, and finite volume method are grid/mesh-based methods often used in solving Stefan problem. Voller and Shadabi [2] used finite difference method via enthalpy formulation for solving two-dimensional Stefan problem. Moreover, the finite element method [3-5] and the finite volume method [6, 7] are widely employed for solving both one-dimensional and two-dimensional Stefan problem.

The smoothed particle hydrodynamics (SPH) method is a meshless method. This method is a particle-based interpolation which was originally used for modelling astrophysical problems. However, this method is eventually developed in fluid dynamics problems. In this paper, we propose SPH method for solving two-dimensional Stefan problem. The SPH method which is applied in melting problem together with fluid flow could be seen in [13]. In other case, Monaghan et al. [8] used SPH method in solving solidification problem, but the fluid flow is not considered in the model.

In the present case, we employ SPH method to simulate ice melting problem in two-dimensional systems. For the reverse problem that is water solidification, the mathematical formulation is the same but different in the initial and boundary condition. In the melting or solidification case, two regions of the system are considered, liquid (water) and solid (ice) separated by phase change interface. Moreover, the interface is always moving in each time step. The system leads us to the two-phase Stefan problem.

Two heat conduction equations should be solved in solid and liquid region, but the boundary of the domain is not known. Our goal is to determine temperature field $T(x,y,t)$ that satisfy a heat conduction equation inside solid region, a heat conduction equation inside liquid region. Also, $T(x,y,t)$ must satisfy interface condition, initial and boundary condition.

## 2. GOVERNING EQUATION

There are three categories for governing equations of the system: fluid flow, solid motion, and heat transfer (energy equation). The fluid motion is assumed weakly compressible. The governing equations for weakly compressible fluid are the Navier-Stokes and the continuity equations as follow:

$$\frac{D\mathbf{v}}{Dt} = -\frac{1}{\rho}\nabla p + \mathbf{F} \quad \text{...............................................................(1)}$$

$$\frac{D\rho}{Dt} = -\rho \nabla \cdot \mathbf{v} \quad \text{.......................................................................(2)}$$

where $\mathbf{v}$, $\rho$, and $p$ are velocity, density and pressure of the fluid, respectively. While $\mathbf{F}$ is external force per unit mass which can be gravitational acceleration or repulsive force per unit mass. For the present case, surface tension is assumed not to be significant, so it can be neglected.

The equation of state in simple form is given by

$$p = c^2 (\rho - \rho_0) + p_{atm} \quad \text{.........................................................(3)}$$

where $c$ is speed of sound, $\rho_0$ is density reference, and $p_{atm}$ is atmospheric pressure. The choice of speed of sound is very important in the simulation since it influences the density variation. In order to keep density fluctuations less than 1%, $c = \sqrt{200gH}$ is chosen where *g* is gravity acceleration and *H* is height of water level.

The governing equation of heat transfer and solid motion is discussed further in Section 3 and Section 6, respectively.

## 3. ENTHALPY METHOD

Enthalpy method is usually used in modelling that involved phase-change such as melting and solidification. This method refers to reformulate the heat conduction equation in different phases into the enthalpy equation. Also, by using enthalpy equation the latent heat (heat of fusion) is accounted in calculating the temperature field. In the melting case, this latent heat is needed to break up the binding in the solid structure. Otherwise, in the solidification case, the latent heat is released.

One of the advantages of the enthalpy method is, the heat conduction equation in solid and liquid region can be solved without need to know the interface position and the heat flux in the interface is automatically continuous. Moreover, if we know the relationship between enthalpy and temperature then we can calculate temperature field. If we assume heat transfer by conduction only, then the enthalpy equation is

$$\frac{\partial H}{\partial t} = \frac{1}{\rho} \nabla \cdot (k \nabla T) \quad \text{.....................................................(4)}$$

where $H$ is enthalpy, $\rho$ is density and $k$ is thermal conductivity. The relationship between enthalpy and temperature is given by [1]

$$T = \begin{cases} T_m + \dfrac{H}{c_s}, & H \leq 0 \quad \text{(solid)} \\ T_m, & 0 < H < L \quad \text{(interface)} \\ T_m + \dfrac{H-L}{c_l}, & H \geq L \quad \text{(liquid)} \end{cases} \quad \text{...............................(5)}$$

$T_m$ is melting point, $c_s$ and $c_l$ are specific heat of solid and liquid respectively, and $L$ is latent heat.

## 4. SPH FORMULATION

The main idea of the SPH method is to represent a function or its derivative into integral representation and discretize it into set of particles. The detail about SPH method is discussed in [14]. The integral interpolant of function $f(\mathbf{x})$ is

$$f(\mathbf{x}) = \int_\Omega f(\mathbf{x}') \delta(\mathbf{x} - \mathbf{x}') \, d\mathbf{x}' \quad \text{......................................(6)}$$

where $\delta(\mathbf{x} - \mathbf{x}')$ is the Diract delta function and $\Omega$ is the volume of the integral that contain $\mathbf{x}$. If we replace the delta function by smoothing function or kernel function, $W$, we get kernel aproximation of function $f(\mathbf{x})$:

$$\langle f(\mathbf{x}) \rangle = \int_\Omega f(\mathbf{x}') W(\mathbf{x} - \mathbf{x}', h) \, d\mathbf{x}' \quad \text{....................................(7)}$$

where $h$ is smoothing length that represents influence area of kernel function. We also can get the kernel aproximation of the derivative of function $f(\mathbf{x})$:

$$\langle \nabla \cdot f(\mathbf{x}) \rangle = \int_\Omega f(\mathbf{x}') \cdot \nabla W(\mathbf{x} - \mathbf{x}', h) \, d\mathbf{x}' \quad \text{....................................(8)}$$

The following equations are derived by using SPH discretization:

(Navier-Stokes equation) $\quad \dfrac{d\mathbf{v}_a}{dt} = -\sum_b m_b \left( \dfrac{p_b + p_a}{\rho_a \rho_b} \right) \nabla_a W(|\mathbf{r}_a - \mathbf{r}_b|, h) \quad \text{......................(9)}$

(continuity equation) $\quad \dfrac{d\rho_a}{dt} = \sum_b m_b (\mathbf{r}_a - \mathbf{r}_b) \cdot \nabla_a W(|\mathbf{r}_a - \mathbf{r}_b|, h) \quad \text{............................(10)}$

Equation (9) and (10) mean that the velocity and density in a particle can be obtained by summing up the contribution of the neighboring particles. The contribution is governed by kernel function $W$. Therefore the choice of $W$ will affect the accuracy of the simulation. There are many choices of kernel function, but for the present simulation, the following cubic spline kernel is used:

$$W(q,h) = \dfrac{10}{7\pi h^2} \begin{cases} 1 - \tfrac{3}{2} q^2 + \tfrac{3}{4} q^3, & 0 \leq q < 1 \\ \tfrac{1}{4}(2-q)^3, & 1 \leq q < 2 \\ 0, & \text{otherwise} \end{cases} \quad \text{............................(11)}$$

where $q = |\mathbf{r}_a - \mathbf{r}_b|/h$.

The heat transfer between particles (water, ice, and solid boundary) is represented by the temperature and enthalpy which are carried by SPH particles. In the present simulation there are only two types of heat transfer to be taken into account of the conduction and convection. But, the heat transfer by convection is automatically fulfilled because the liquid particle is always moves according to the Navier-Stokes equation [13]. The heat transfer by conduction in SPH form is given by [9]

$$\frac{dH_a}{dt} = \sum_b \frac{m_b}{\rho_a \rho_b} \frac{k_a k_b}{k_a + k_b}(T_a - T_b)\frac{(\mathbf{r}_a - \mathbf{r}_b)\cdot\nabla_a W(|\mathbf{r}_a - \mathbf{r}_b|,h)}{|\mathbf{r}_a - \mathbf{r}_b|^2 + \eta^2} \quad\quad (12)$$

where $\eta = 0.01h$.

The steps in constructing the SPH heat conduction of equation (12) is briefly discussed in [11]. Note that by using equation (12), in interface between solid and liquid phase, the heat flux is automatically continuous. Once we obtain the enthalpy of the particles, the temperature particles and phase-change transition are determined by using equation (5). The leap-frog time stepping is employed for equation (9) and equation (10) while explicit Euler time stepping is used for equation (12).

## 5. BOUNDARY TREATMENT

In SPH modelling, it is very important to treat the particles that approach wall boundaries. In this simulation, the wall boundaries are assumed as solid boundary particles. We need these solid boundary particles to prevent the water particles penetrate the wall boundaries. The water particles approaching wall boundaries will experience repulsive force coming from solid particles. The force is experienced by a water particle $j$ normal to the solid boundary particle $i$ is given by [10]

$$\mathbf{f}_j = \mathbf{n}_i R(y) P(x) \quad\quad (14)$$

where $\mathbf{n}_i$ is normal vector of the solid boundary. Here,

$$R(y) = A\frac{1}{\sqrt{q}}(1-q) \text{ and } P(x) = \tfrac{1}{2}\left(1+\cos\left(\tfrac{\pi x}{\Delta p}\right)\right)$$

where $x$ is distance of projection water particle $j$ on tangent vector of the solid boundary particle $i$, $y$ is perpendicular distance of water particle $j$ from the solid boundary particle $i$ (see Figure 1), $q = y/(2\Delta p)$ and $\Delta p$ is initial particle spacing. The choice of particle spacing of solid boundary particles depends on the initial water particle spacing. If $\Delta p$ is too small, then the water particle will experience a repulsive force before it approaches the wall boundary. Whereas if $\Delta p$ is too big, some of water particles may penetrate the boundary wall before experiencing the repulsive force.

## 6. FLOATING ICE MODEL

There are two ways for modeling ice (solid) objects. First, consider ice particles as if they were water particles. Hence, both solid and liquid particles are solved by using Navier-Stokes equation. In this way the density of ice particles is also calculated by using equation (2). But, they are not updated in each time step in order to keep the consistency of the density of ice. However, it makes problems because ice density is 917 kg/m$^3$ and water density is 1000 kg/m$^3$ (the difference is too wide) while the Navier-Stokes equation is designed for weakly compressible. Consequently, ice particles will have low pressure, and it makes water particles penetrate ice objects. The second way is to consider ice particles as solid boundary (wall) particles. In this way, If the water particles approaching ice particles, they will experience repulsive force. The problem is we have to find normal vector of ice objects in each time step. It is slightly difficult since ice formation is always changing. In melting case, it may be easy to find normal vector of ice wall because at initial we set ice particles neatly. But, in solidification case, it is rather difficult to find normal vector of ice wall which is formed from water particles. It is because the ice particles formed are unstructured.

In the present simulation, the ice objects which float in water are assumed as wall boundary particles. So that for each water particle that approaches the ice particles, it will experience repulsive force which is calculated according to equation (14). But, the challenge is in determining the normal vector of ice wall since the wall of ice is changing and moving in each time step. The normal vector of ice particle $i$ due to water particle $j$ can be determined by projecting vector $\mathbf{r}_{ij}$ onto the tangent vector of particle $i$. To find tangent vector, it is necessary to find two neighboring particles of particle $i$ says $i$-1 and $i$+1 with the opposite position, see Figure 1. The tangent vector $\mathbf{t}_i$ is $\mathbf{r}_{i+1} - \mathbf{r}_{i-1}$. Hence the unit normal vector is

$$\mathbf{n}_i = (\mathbf{r}_{ij} - \mathbf{proj}_{\mathbf{t}_i}\mathbf{r}_{ij})/\left|\mathbf{r}_{ij} - \mathbf{proj}_{\mathbf{t}_i}\mathbf{r}_{ij}\right|$$

where $\mathbf{proj}_{\mathbf{t}_i}\mathbf{r}_{ij}$ is projection of vector $\mathbf{r}_{ij}$ onto tangent vector $\mathbf{t}_i$. The advantage of this strategy is that $\mathbf{n}_i$ is always outward normal vector.

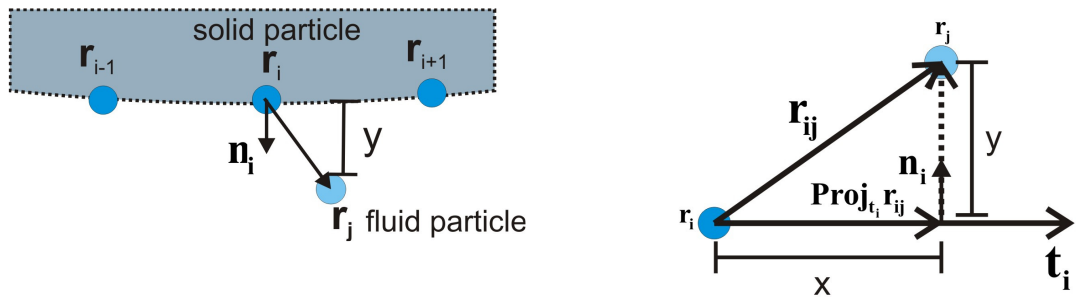

Figure 1. Normal vector and tangent vector

The repulsive force of solid particle $i$ is calculated by summing up the contribution of surrounding water particles [15]

$$\mathbf{f}_i = \sum_{j \in WPs} \mathbf{f}_{ij} \quad \dots\dots\dots\dots\dots\dots\dots\dots\dots\dots\dots\dots\dots\dots\dots(15)$$

where *WPs* denotes water particles. The equations of motion for ice objects are derived from the basic rigid dynamics equation. The translational and rotational equations of solid motion are

$$M\frac{d\mathbf{V}}{dt} = \sum_{i \in IPs} m_i \mathbf{f}_i \quad \text{and} \quad I\frac{d\mathbf{\Omega}}{dt} = \sum_{i \in IPs} m_i (\mathbf{r}_i - \mathbf{R}_0) \times \mathbf{f}_i \quad \dots\dots\dots\dots\dots\dots\dots(16)$$

where $M$, $I$, $\mathbf{V}$, $\mathbf{\Omega}$, and $\mathbf{R}_0$ are mass, moment of inertia, velocity, rotational velocity, and center of mass of the ice objects, respectively. Here, *IPs* represents ice particles. The velocity for each ice particle is given by

$$\frac{d\mathbf{r}_i}{dt} = \mathbf{V} + \mathbf{\Omega} \times (\mathbf{r}_i - \mathbf{R}_0) \quad \ldots\ldots\ldots\ldots\ldots\ldots\ldots\ldots\ldots\ldots\ldots\ldots\ldots\ldots\ldots..(17)$$

Thus the position of ice particles for each time step can be calculated by integrating equation (17) in time.

An ice object may be separated into some ice objects during melting process and also new ice objects may be formed during solidification process. To control the motion of these objects, there should be a strategy to separate the ice particles into certain ice objects, so that equation (16) can be employed. A simple strategy to separate the solid particles has been proposed in [12]. First, each ice particle is assigned an index from 1 to *N*, where *N* is the number of ice particles. Second, each ice particle index is updated iteratively by the maximum index of the neighboring particles. Here, the neighbor of a particle is defined as all particles with distance less than or equal to certain number. The last, the ice particles index that belong to the same object will converge to the maximum index of the ice particles in that objects.

## 7. RESULTS AND DISCUSSION

### 7.1 Physical setup of the ice melting

The physical setup of the system in 2D can be described as follow. Consider a water tank and two ice cubes with the dimension of each ice cube is 0.125 m x 0.125 m. The dimension of water tank is 1 m x 1 m, and at first filled with water of height 0.25 m as seen in Figure 2. The initial temperature of the first and the second ice cube are -15 $^0C$ and -10 $^0C$, respectively, and the temperature of water is 30 $^0C$. The outside boundaries of the system are adiabatic with the initial temperature is -10 $^0C$. The heat transfer in the system is assumed to occur only by conduction and convection. The 11322 particles are used for this simulation. All parameters for this simulation can be seen in Table 1. For solidification case, the physical setup is generally same but different in initial and boundary condition. The simulation result for solidification case may be shown later in other paper.

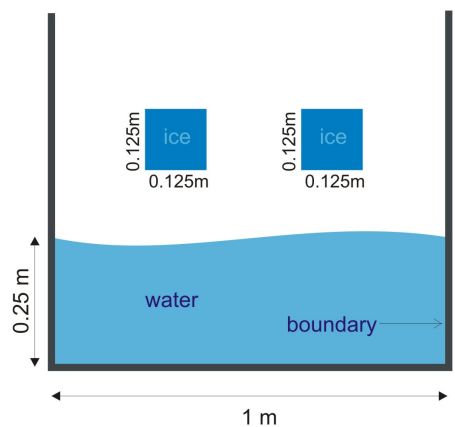

Figure 2. Physical setup of the ice melting

Table 1. Parameter for the simulation

| No | Parameter | Value | Unit |
|---|---|---|---|
| 1 | Specific heat of ice | 2.067 | kJ/kg$^0$C |
| 2 | Specific heat of water | 4.219 | kJ/kg$^0$C |

| 3 | Conductivity of ice | 2.216e-3 | kJ/ms$^0$C |
|---|---|---|---|
| 4 | Conductivity of water | 0.566e-3 | kJ/ms$^0$C |
| 5 | Melting point | 0 | $^0$C |
| 6 | Latent heat | 334 | kJ/kg |
| 7 | Density of ice | 917 | kg/m$^3$ |
| 8 | Density of water | 1000 | kg/m$^3$ |

**7.2 Simulation results of ice melting**

At first, two ice cubes fall freely from height of 0.25 m above water surface. These two ice cubes then interact with the water. The forces acting on the water particles and ice particles are calculated by using equation (1) and equation (16) respectively. To get temperature of the particles, heat transfer in each particle is determined by the enthalpy equation (4). Temperature distribution of both solid particles and liquid particles for several time steps can be seen in Figure 3. In that figure, the color represents the temperature of particles in certain position.

Each image in Figure 3 represents not only the phase change from solid to liquid, but also the patterns of convection flow. The convection patterns are quite interesting to be explored further. These patterns are caused by differences in density between new water particles and old water particles. The new water particles mean ice particles which have just transformed to water particles. In this simulation, the ice particles own have fixed density that are 917 kg/m$^3$, while the water particles have density in the range of 998-1005 kg/m$^3$ (weakly compressible fluid). However, after the ice particles change the phase from solid to liquid, the density of the new water particles are changed to 1000 kg/m$^3$. The new water particles caused density difference quite striking with the surrounding water particles. There will be movement of the particles according to equation (1) and (2).

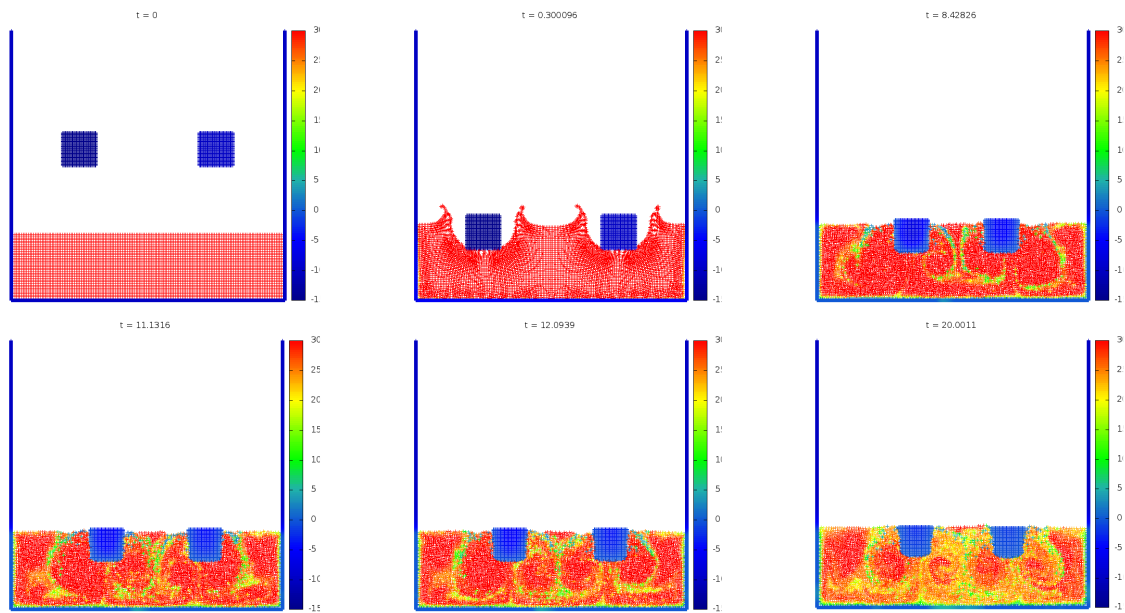

Figure 3. Temperature distribution for several time steps

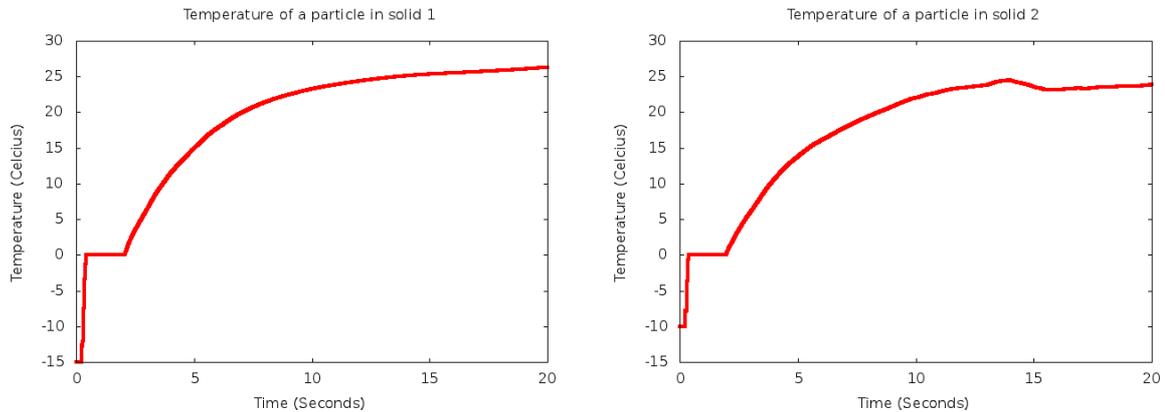

Figure 4. Temperature vs time of a single particle of the first and second ice cube

The left graphic in Figure 4 shows the temperature of a single particle that initially located on the edge of the first ice cubes. The initial temperature of the particle is -15 $^0C$, then the temperature is gradually changed to 0 $^0C$ in 0. 426 seconds. However, in this condition, the state is still in the solid phase. To change the phase of this particle from solid to liquid the energy is needed for amount latent heat (heat of fusion). The latent heat per particle is $m_0L$ = 2.99 kJ. After the particle state is changed into liquid the temperature will be influenced by the temperature of the surrounding water particles. Total time required for a single particle in first ice cube to change the phase from solid to liquid is about 2.06 seconds. While the right graphic of Figure 4 shows the temperature of a single particle of the second ice cube that has the initial temperature of -10 $^0C$. For a single particle in the second ice cube, the total time needed to change the phase is about 1.96 seconds. Total time that required by an ice particle to melt depends on the mass and latent heat of the particle.

## 6. CONCLUSION

In this paper we have shown that an SPH method can be used to simulate Stefan problem especially in ice melting problem. The advantage of the SPH method is all physical information such as density, pressure, velocity, and temperature can be stored in an SPH particle. With this way, it is easy to treat the interaction between solid and liquid phase which is very hard if on the contrary, the mesh-based method is used. By using enthalpy method via SPH formulation, heat conduction equation in solid and liquid region can be solved without needed to know the interface position between solid and liquid. As though, this model is new in SPH development, so it needs many improvements to have more realistic simulation.

**Acknowledgements**
The author would like to thank Prof. Seiro Omata for giving very helpful comments and suggestions. The author also thanks Dr. Masaki Kazama for the useful discussions.

**REFERENCES**
[1] Alexiades, V. and Solomon, A. D., 1993, *Mathematical Modeling of Melting and Freezing Processes*, Taylor & Francis, Washington, DC.

[2] Voller, V.R. and Shadabi, L., 1984, Enthalpy methods for tracking a phase change boundary in two dimensions, *Int. Comm. Heat and Mass Transfer*, **11**, 3, pp. 239-249.
[3] Beckett, G. and Mackenzie, J.A. and Robertson, M.L., 2001, A Moving Mesh Finite Element Method for the Solution of Two-Dimensional Stefan Problems, *J. Comput. Phys.*, **168**, 2, pp. 500-518.
[4] Bonnerot, R. and Jamet, P., 1977, Numerical computation of the free boundary for the two-dimensional Stefan problem by space-time finite elements, *J. Comput. Phys.*, **25**, 2, pp. 163-181.
[5] Salvatori, L. and Tosi, N., 2009, Stefan Problem through Extended Finite Elements: Review and Further Investigations, *Algorithms*, **2**, 3, pp. 1177-1220.
[6] Lan, C.W. and Liu, C.C. and Hsu, C.M., 2002, An Adaptive Finite Volume Method for Incompressible Heat Flow Problems in Solidification, *J. Comput. Phys.*, **178**, 2, pp. 464-497.
[7] Voller, V. R. and Swaminathan, C. R. and Thomas, B. G., 1990, Fixed grid techniques for phase change problems: A review, *Int. J. Num. Methods in Engineering*, **30**, 4, pp. 875-898.
[8] Monaghan, J.J. and Huppert, H. E. and Worster, M.G., 2005, Solidification using smoothed particle hydrodynamics, *J. Comput. Phys.*, **206**, 2, pp. 684-705.
[9] Cleary, P.W., 1998, Modelling confined multi-material heat and mass flows using SPH, *Applied Mathematical Modelling*, **22**, 12, pp. 981-993.
[10] Monaghan, J.J. and Kos, A., 1999, Solitary waves on Cretan beach, *J. Waterway, Port, Coastal, Ocean Eng*, **125**, 3, pp. 145-154.
[11] Cleary, P.W. and Monaghan, J.J., 1999, Conduction Modelling Using Smoothed Particle Hydrodynamics, *J. Comput. Phys.*, **148**, 1, pp. 227-264.
[12] Iwasaki, K. and Uchida, H. and Dobashi, Y. and Nishita, T., 2010, Fast Particle-based Visual Simulation of Ice Melting, *Computer Graphics Forum*, **29**, 7, pp. 2215-2223.
[13] Tong, M. and Brown, D.J., 2012. Smoothed particle hydrodynamics modelling of the fluid flow and heat transfer in the weld pool during laser spot welding , *OP Conf. Ser.: Mater. Sci.*, **27**.
[14] Liu, M. and Liu, G., 2010, Smoothed Particle Hydrodynamics (SPH): an Overview and Recent Developments, *Archives of Computational Methods in Engineering*, 17, 1, pp. 25 – 76.
[15] Monaghan, J.J., 2005, Smoothed particle hydrodynamics, *Rep. Prog. Physc.*, **68**, 8, pp. 1703-1759.